\documentstyle[twocolumn,eqsecnum,aps,epsfig]{revtex}

\def\be{\begin{equation}}
\def\ee{\end{equation}}
\def\bea{\begin{eqnarray}}
\def\eea{\end{eqnarray}}

\begin{document}

\title{Highly Sensitive  Centrality Dependence of Elliptic Flow -- 
         A Novel Signature of the Phase Transition in QCD   }
\author{H. Sorge}    
\address{\em  Department of Physics \& Astronomy, 
                         SUNY at Stony Brook, NY 11794, USA} 
                                     
\date{December 21st, 1998}
\maketitle   
\begin{abstract}
Elliptic flow of the hot, dense system which has been created in nucleus-nucleus
collisions develops as a response  to the initial azimuthal asymmetry of the
reaction region. Here it is suggested that the magnitude  of this response shows
a ``kinky'' dependence on the centrality of  collisions for which the system passes
through a first-order or rapid transition between quark-gluon plasma and
hadronic matter. We have studied the  system Pb(158AGeV)  on Pb employing a
recent version of the transport theoretical approach RQMD and find the
conjecture confirmed. The novel phase transition signature may be observable in
present and forthcoming experiments at CERN-SPS and at RHIC, the BNL collider.  
\end{abstract}
%
%
\narrowtext

One of the most important goals of the heavy-ion programs
in the ultrarelativistic energy domain is the search for the
phase transition between hadronic matter and the quark-gluon plasma
(QGP). After a decade-long effort based on numerous experiments
 at fixed-target machines (CERN-SPS, BNL-AGS)  heavy-ion physics
can be considered a mature field today. Thus it may seem surprising
that there still is a shortage of reliable signatures
for  the elusive state  QGP and the transition itself \cite{BGGS98}.
It may be easier to comprehend the 
difficulties to identify the QGP if one takes into account that
properties of the QGP and the transition have to be reconstructed
from the final state which obviously is of  hadronic nature.
On a more fundamental level, it has become clear only recently that
the properties of strongly interacting matter even far above the
critical temperature $T_c$ are essentially
non-perturbative. 
This makes many of
the ``first-generation'' QGP signals which are based on
perturbation theory unreliable at best.   
 
On the other side,   information 
about the  QGP and the phase transition region has become available
with the advent of more powerful       
lattice gauge simulations  of quantum chromodynamics (QCD)
\cite{lat1,lat2}. Most notably, it has been
shown that chiral symmetry is restored at rather low temperatures
(in the range 140 to 170 MeV). Furthermore, the Equation of State (EOS)
varies rather rapidly in the transition region. It is not clear yet
whether the transition is of weak first-order  or  just a rapid
cross-over between the two phases. 
The EOS extracted from the lattice clearly displays the transition 
from hadron to quark-gluon degrees of freedoms. 
Pressure and energy density approach the ideal
Stefan-Boltzmann values at temperatures $\ge 3 T_c$.
A generic feature of the EOS in the transition  region is the
presence of the so-called ``softest point of the EOS'' 
\cite{MHS95,RPM95}
related to the effect that the energy density may jump with
increasing temperature but the pressure  does not.
 
The collective transverse flow which
develops in the heavy-ion collisions reflects on  the properties
of the EOS. 
Usually, one distinguishes various types of transverse flow,
the radial (isotropic component), directed (sideward kick
in the reaction plane) and the elliptic flow, the latter being
a preferential emission either along the impact parameter axis
or out of the reaction plane (squeeze-out) \cite{PV98}.
The general idea why a phase transition may show up in flow observables
is rather straightforward. 
At densities around the softest point the
 force driving the matter expansion gets weakened. 
A long time ago, van Hove has suggested that the multiplicity
dependence of  average transverse momenta 
may display a plateau and a second rise \cite{vH82}. 
So far, it has not been possible to deduce
the presence of a phase transition  from the 
transverse momentum spectra.  
Some time ago we have suggested that
the elliptic  flow may be a                 
better-suited observable to identify a first-order type phase transition  
\cite{HSplb97}. Here we make good on this promise and present
a novel signature of the QCD phase transition.  We
 predict a
rather characteristic centrality dependence of the elliptic flow 
if the created system passes through the softest region of the EOS in the
heavy-ion reactions. 
                     
Elliptic flow in the central region of ultrarelativistic
collisions is driven by the almond-shape of the participant matter
in the transverse plane \cite{ollitrault}.
It was argued in \cite{HSplb97} that elliptic flow may be more
sensitive to the early pressure than the isotropic radial flow.
``Early'' and ``late'' is defined by the time scale set by
the initial transverse size 
 $r_t=\sqrt{\langle x^2 +y^2 \rangle}$ 
 of the reaction region. 
 Thus early flow 
 appears at times  $\approx r_t/c $ while we would refer to
flow generated at times $>2r_t/c$ as late. 
One reason for the larger sensitivity of the elliptic flow to early
pressure is that  the generated flow asymmetry  works against
its cause and diminishes the spatial asymmetry on a time scale proportional to
  $\sqrt{\langle y^2 \rangle}-\sqrt{\langle x^2 \rangle}$. 
Furthermore, the elliptic asymmetry  is proportional to the difference
 between the flow strengths in  $x$ 
  (parallel to impact parameter) and $y$ direction. Thus it is 
  more fragile than radial flow.  Viscosity  related non-ideal effects
  tend to wash out the pressure-driven asymmetries. 
   Obviously, these effects will be  more pronounced in the later
     dilute stages of the reaction.
  Unfortunately, this  could not
 be demonstrated in the earlier work. The transport model   RQMD 
 (version 2.3) \cite{SOR95}  employed for the calculations lacked any sizable 
 transverse pressure in the early stages --       
   a combination of softness from pre-equilibrium motion    
    and absence of a QGP phase which would generate more pressure
     than the resonance matter  simulated in the model.
   As a result the final hadron momentum spectra
   showed  azimuthal asymmetries much smaller than hydrodynamical
   results which include a phase transition into the QGP. 
 In the mean time, NA49 has analysed  data
   for semi-central Pb(158AGeV) on Pb collisions \cite{NA49}.   
    The measured 
   azimuthal asymmetries 
     are roughly equally distant from the closest results 
     based on  hydrodynamics
     and from the RQMD calculations  \cite{olli98,LPX98}. 
      Both of these calculations  show a
   factor of two disagreement, however, in different directions.     
  In this Letter we are going to present results from calculations
   with a new version of the transport model  RQMD (version 3.0)
   which incorporates an EOS with 1st order phase transition.
   Comparing to the results obtained in the model without QGP phase
   we may assess the importance of the phase transition. 
\vspace{-1.cm}
\begin{center}
\begin{figure}[h]
\centerline{\epsfig{figure=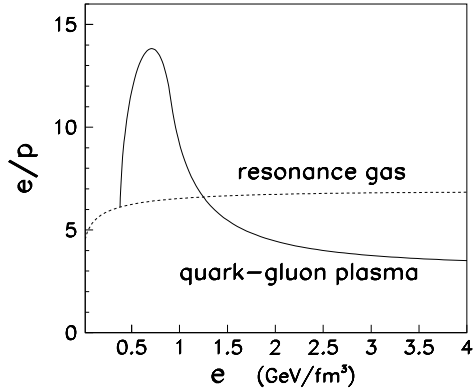,width=7.0cm}}
\caption{
 Equation of states implemented into RQMD:
 ratio of energy density $e$ divided by pressure $p$.
 The dashed line  represents the resonance gas EOS,
 the solid line  the EOS including a first order
  phase transition
    with $T_c$=160 MeV.
 }
\end{figure}
\end{center}
\vspace{-0.5cm}     
 
 Let us first shortly describe how the phase transition is implemented
 into the model. A detailed description of the algorithm
  will be presented elsewhere. In RQMD 
  nucleus-nucleus collisions are calculated in a Monte Carlo type fashion.
  While the nucleons from each colliding nuclei pass through each other,
   they are decomposed into constituent quarks.
  Strings may be excited, and overlapping strings fuse
   into ropes (with larger chromoelectric field strength).
    After their decay and fragmentation  secondaries emerge and may interact
   with each other. Formed resonances are treated as unstable 
   quasi-particles.  This  leads to a resonance gas    EOS 
   if there are no corrections  from other interactions.    
  The QCD dynamics in the phase transition region is not well understood.
   Even if there is a quasi-particle description it is not obvious
    which one  of the many possible choices (strings, constituent quarks,
    partons, either  massless or with dynamical masses) 
    is to be prefered.    In this situation  we  have decided to
   stick to the implemented degrees of freedom and modify the collision
   term instead. Since we expect that hydrodynamics is a reasonable
    approach  for the transverse dynamics in the ultradense stage,
    the EOS should be the most relevant ingredient 
     for the expansion dynamics anyway. 
  It is well-known that different treatment of
   interactions between quasi-particles may modify the EOS. In general,
    if particles are free between interaction points   the virial
    theorem specifies that the pressure of the system in equilibrium
   is given by  \cite{DP96}
 \begin{eqnarray}
   \label{virialeq}
   P &= &
            P_{id} +
        \frac{1}{d\cdot V\cdot \Delta T}
          \sum _{(a,b)}
          \left (
            \delta \vec{p} _{a} \cdot \vec{r}_{a}
            + \delta \vec{p} _{b} \cdot \vec{r}_{b}
             \right )
         \quad .
 \end{eqnarray}
 The first term arises from free streaming. The second term
  represents the non-ideal contribution $\Delta P $
  due to momentum changes $\delta \vec{p} _{a}$ 
   at discrete collision  points $\vec{r}_{a}$.
   $d$=3 is the number of space dimensions, $V$ the volume of the system,
  $\Delta T $ a sufficently large time interval, and the summation goes
   over all collisions.
   $a$ and $b$  specifies which quasi-particles collide.
  The standard collision term in  RQMD  is manufactured such
   that it does not contribute to the pressure. Now, we 
   depart from this ``ideal'' collision term and
    let each quasi-particle interact elastically with a neighbor
   after any of the standard collisions. 
   The momentum change is constrained  by 
 \begin{eqnarray}
   \label{dpreq}
 \left \langle \delta \vec{p} _{a} \vec{r}_{a} \right \rangle
 +\left \langle \delta \vec{p} _{b} \vec{r}_{b} \right \rangle
    &\stackrel{!}{=}& d \cdot \frac{\Delta P}{\rho}
         \cdot \left (    \Delta t _a ^{sc}
                  + \Delta t _b ^{sc}  \right )
    \quad . 
 \end{eqnarray}
 $\Delta t _a ^{sc} $ refers to the time which has elapsed since the
 last of the EOS modulating collisions. $\rho$ is the density of
  quasi-particles. Introducing collisions according to eq.~(\ref{dpreq})
  changes the pressure of the system to  $P_{id}+\Delta P$.
 Eq.~(\ref{dpreq}) provides a numerically rather efficient 
 method to modify  the ideal EOS. 
  The physics content of eq.~(\ref{dpreq}) is that the momentum
   transfer may be chosen to be either suitably repulsive 
   (QGP at high temperature) or attractive
    (mixed phase).
  Fig.~1 displays the ideal EOS based on  counting  the
  propagating quasi-particles in RQMD. In addition, an  EOS
  is shown which may be produced by  introducing 
  energy density dependent  additional interactions according to 
  eq.~(\ref{dpreq}). This EOS is the one which will be used for the 
  calculations presented in this Letter. It is calculated  from a
  quasi-particle model  of quarks and gluons with dynamical thermal
  masses  \cite{P96,K97}.
  We have chosen this EOS, because it provides a good fit to lattice data.
  The EOS contains a 1st order transition at $T_c$=160MeV with a latent heat
   of 467 MeV/fm$^3$.
  For the RQMD calculations of nucleus-nucleus
  collisions the novel interaction term is introduced in a local
  density approximation, i.e.\ all variables in eq.~(\ref{dpreq})
   are evaluated in the local rest system of the energy current.
  Neither is  the modulation of the local pressure tensor  restricted
  to regions of  local equilibrium nor is 
   --  the other extreme -- any  local equilibration enforced,
   e.g. by randomizing  directions of local momenta. 
\vspace{-1.cm}
\begin{center}
\begin{figure}[h]
\centerline{\epsfig{figure=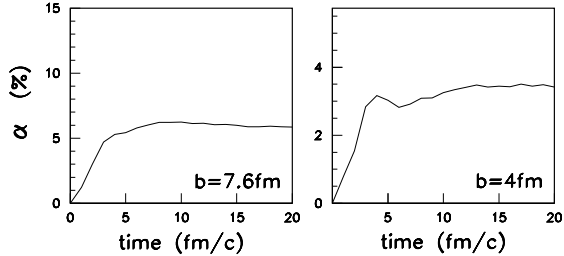,width=7.0cm}}
\caption{
 Time evolution of transverse momentum anisotropy parameter
   $\alpha $ calculated with RQMD for the  system
   Pb(158AGeV) on Pb at two impact parameters.
  Only quasi-particles within a central rapidity cut ($\pm 0.7$)
   are included.
 }
\end{figure}
\end{center}
\vspace{-0.5cm}     
  
Let us now turn to results  of RQMD calculations which contain a
 phase transition. 
 We have chosen to do the calculation for the system  Pb(158AGeV) on Pb,
 i.e.\ the heavy-ion reaction  at highest  beam energy which is
   currently   available. 
This may be a good place to look for the  phase transition. 
 The time evolution of the azimuthal asymmetry parameter
  $\alpha $ (momentum flow asymmetry)
  \begin{equation}
     \alpha =
          \left(
             \langle  p_x^2  \rangle
            -  \langle  p_y^2  \rangle
          \right)/
          \left(
             \langle  p_x^2  \rangle
            +  \langle  p_y^2  \rangle
          \right)
   \label{eqalpha}
  \end{equation}   
 for quasi-particles around midrapidity is displayed in fig.~2. 
 It shows a very different behaviour than corresponding calculations
  based on RQMD without QGP-type EOS \cite{HSplb97}. 
 We see from the figure that in case of QGP formation essentially
 all of the finally observable asymmetry is created at times smaller
  than 4~fm/c. The mixed phase leads to a marked dip of the asymmetry
  for more central collisions. Since the pressure is comparably low,
  free motion between interactions is able to destroy some of the 
  earlier created  flow asymmetry.  
  On the other side, the calculations for semi-peripheral collisions
   (e.g. b=7.6 fm) show that the softening in the mixed phase
   cannot stall the expansion of the system. Needless to say that
   this is a function of the latent heat which is very   moderate 
   for the chosen EOS.
  The overall effect of mixed phase and purely hadronic stage is
  rather small in a broad  impact parameter range.
   Under the condition of an already well-developed flow asymmetry, 
   diffusive processes and thermal pressure driven work 
   seem to neutralize each other at the later stages. 
   In the QGP-scenario the azimuthal asymmetry
  is indeed mostly a signature of the early pressure. 
   It is amuzing that  non-ideal effects from viscosity in the
   low-density stage
   may be helpful to infer information about the pressure in the
   high-density region.

In the following we will present the main result of the Letter,
the measurable azimuthal asymmetry of final hadrons 
 which the experimentalists usually take
to be the number flow asymmetry $v_2 $
\be
  v_2= \langle  \cos (2\phi )  \rangle
\ee
as a function of centrality. Tight impact parameter cuts can be obtained
using a forward-energy trigger like NA49 has. 
Of course, the spatial asymmetry of the reaction zone which is 
correlated with the asymmetry of the participant nucleons in the ingoing nuclei
\be
     \alpha _x=
          \left(
             \langle  y^2  \rangle
            -  \langle  x^2  \rangle
          \right)/
          \left(
             \langle  x^2  \rangle
            +  \langle  y^2  \rangle
          \right)
         \qquad .            
\ee
is itself a function of the impact parameter. Trivially,  $v_2$  goes
to zero for  very small and very large impact parameters. 
 The value for $v_2$ at any given centrality reflects both the strength of
the spatial asymmetry and the response of the created system due to the
generated pressure.
However, we may disentangle the effects from geometry and 
dynamics. In general, the final flow asymmetry
$v_2$ can be viewed as a function of many variables, 
 $\alpha_x$, the 
average initial energy density  $e_0$, 
 the transverse size $r_t$, to name just a few:
\be
 v_2 =  f(\alpha_x, e_0, r_t, ...) 
     \approx  A_2(\overline{\alpha_x}) \cdot \alpha_x
                 + {\cal {O}}((\alpha_x-\overline{\alpha_x})^2)
     \quad ,
\ee
where we have obtained the second equation from a Taylor expansion around
some intermediate value $\overline{\alpha_x}$ and taking into
account  that $v_2$ vanishes for $\alpha_x \rightarrow 0$.           
 In Pb on Pb collisions,
${\alpha_x}$ varies between 0 and 0.50 for impact parameters 
less than 12 fm.                                    
Picking an intermediate value of 
$\overline{\alpha_x}$ means that the neglected 
higher order
terms in $(\alpha_x-\overline{\alpha_x})$ are expected to
be rather small in practice, 
on the order of
10 percent. Defining the  scaled flow asymmetry as
\be
 A_2 = v_2 /  \alpha_x 
\ee
will therefore allow to
assess the dynamical response of the created system to the  
initial spatial asymmetry.

We display the scaled flow asymmetry $A_2$      
versus impact parameter $b$ in fig.~3.  
Of course,  the asymmetry factor   $A_2$  will tend to vanish in the most
peripheral collisions ($b \approx 2 R_{Pb}$). 
Initial energy densities are too small, and
the system size does not sustain extended reaction times. 
Both for pions and for protons, $A_2$ 
shows a pronounced variation  for  smaller $b$ values.   
This is a result of the EOS softness at intermediate energy densities.
However, non-equilibrium effects, in particular partial thermalization
initially and system-size dependent freeze-out, also play a major role. 
Extracted   $A_2$ values from 
hydrodynamic  calculations  \cite{ollitrault}  show  essentially
no centrality dependence, except for the grazing collisions. 
This feature is in marked contrast to the
transport calculation which includes the  non-equilibrium aspects
of the dynamics.
 
Without phase transition, the asymmetry factor   $A_2$ calculated from
RQMD would simply increase monotoneously 
with centrality -- approximately linearly with the initial system size in the
reaction plane ($\sim 2 R_{Pb}-b$). 
Indeed, 
the hard QGP stage of the reaction leads to a rapid increase
of the asymmetry $A_2$ in collisions with $b\ge $10~fm
as is visible from fig.~3. 
In this range of centralities the initial source size
 $\sqrt{\langle x^2 \rangle }$ along the impact parameter axis is
 small enough that the associated characteristic time for the
 development of  flow falls within the life time of the QGP phase.
In  somewhat more central collisions
further increase of the asymmetry is cut-off after the system enters into the
stage  of soft and lateron viscous expansion.
    Initial energy densities
  change less  with  increasing centrality than the system size. 
 Therefore at the characteristic time for flow development
  typical energy
  densities  are
   in the region of the softest point. In these reactions
 increasing reaction time which is helpful to develop the asymmetry
  is counteracted  by the softness
  of the matter.  In any case, the centrality dependence of the
  flow asymmetry follows a different slope than for the class of
  more peripheral collisions.
  For  collisions  with $b<$ 5~fm
  kinetic equilibration which takes place on a scale of 3-4 fm/c  may
  be realized already in the QGP phase. This gives rise to yet another
    centrality dependence of 
   the flow asymmetry $A_2$ (a second rise).               
\vspace{-1.cm}
\begin{center}
\begin{figure}[h]
\centerline{\epsfig{figure=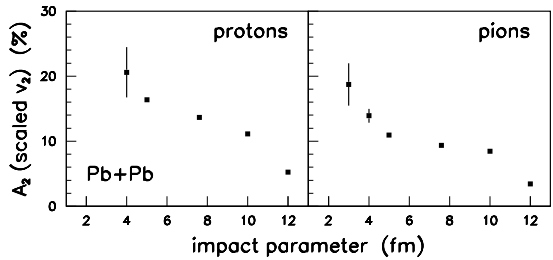,width=7.0cm}}
\caption{
Scaled azimuthal asymmetry parameter $A_2$=$v_2/\alpha_x$ of protons and pions
 as a function of
 impact parameter for the same system and acceptance window as in fig.~2.
 }
\end{figure}
\end{center}
\vspace{-0.5cm}      

  van Hove's original idea to look for  a plateau and a second rise
  in momentum spectra as a signal of  the QCD phase transition 
  may turn out to be true, after all. Present experience tells that
   it will probably not been found in the  multiplicity dependence of
    average transverse momenta. He did not take into account that the dynamics
    of the hadronic stages may add a late radial flow component  \cite{vHSX98}
   which spoils the  original idea.  However, if the presented calculations
   contain some truth it is much better justified
   to neglect the late hadronic stages for the azimuthal asymmetries
    of particle spectra.  
  The presented calculation contains some uncertainties. The
   equation of state is not well-determined in the transition region.
   The admixture of baryons in the central region and the strong
   pre-equilibrium deformation of the local stress tensor add to the
   uncertainties. Nevertheless, the potential rewards  in terms of insight into
    the phase transition dynamics should justify
   a careful search for structure in the centrality dependence of
   elliptic flow at SPS and future RHIC energy.
    
This work has been
supported by DOE grant No. DE-FG02-88ER40388. 
After completion of the manuscript the author became aware  of a  
recent work \cite{HHL98} in which the influence of a 1st order transition on
elliptic flow is also being discussed.

\end{document}